\documentclass[11pt,english]{article}

\usepackage{natbib}
\usepackage{url}
\makeatletter
\def\url@leostyle{%
    \def\UrlFont{\sf}}{\def\UrlFont{\small\ttfamily}}
\makeatother
\usepackage[hidelinks]{hyperref}

\usepackage{authblk}

\usepackage{graphicx}
\usepackage{babel}
\usepackage{amsmath}
\usepackage{ stmaryrd }
\usepackage[titletoc,title]{appendix}


\usepackage{geometry}
\raggedright

\usepackage[T1]{fontenc}

\usepackage{txfonts}
\newcommand{\comment}[1]{}

\usepackage{setspace}
\usepackage[]{footmisc}

\usepackage[font=normalsize,font=doublespacing,labelfont=bf]{caption}

\usepackage{titlesec}
\usepackage{color}
\usepackage[titletoc,title]{appendix}
\titleformat*{\section}{\bf}
\titleformat*{\subsection}{\it}
\titleformat*{\subsubsection}{\it}

\begin{document}

\title{No-Go Theorems:  \\ What Are They Good For?\footnote{Forthcoming in  \textit{Studies in History and Philosophy of Science}: \url{https://doi.org/10.1016/j.shpsa.2021.01.005}}}
\author{Radin Dardashti}
\affil{Interdisciplinary Centre for Science and Technology Studies (IZWT), \\ University of Wuppertal \\ dardashti@uni-wuppertal.de }

\date{}

\maketitle

\thispagestyle{empty}

\begin{abstract}
No-go theorems have played an important role in the development and assessment of scientific theories. They have stopped whole research programs and have given rise to strong ontological commitments. Given the importance they obviously have had in physics and philosophy of physics and the huge amount of literature on the consequences of specific no-go theorems, there has been relatively little attention to the more abstract assessment of no-go theorems as a tool in theory development. We will here provide this abstract assessment of no-go theorems and conclude that the methodological implications one may draw from no-go theorems are in disagreement with the implications that have often been drawn from them in the history of science.

\end{abstract}
\pagebreak

\tableofcontents

\vspace{1em}

\section{Introduction}

The mathematician and polymath John von  Neumann claimed to have proven in his classic \citep{neumann1932mathematische} the impossibility to complete quantum mechanics by hidden variables. A couple of years later, Grete \citet[P. 251]{hermann2017natural} challenges von Neumann, claiming that he ``introduces into its formal assumptions, without justification, a statement equivalent to the thesis to be proven''. Thirty years later, \citet{jauch1963can} state, unaware of Hermann's claim, that ``[t]he question concerning the existence of such hidden variables received an early and rather decisive answer in the form of von Neumann's proof on the mathematical impossibility of such variables in quantum theory''. Three years later \citet{bell1966problem} shows in his seminal work that ``the formal proof of von Neumann does not justify his informal conclusion'', saying later in an interview that  ``the von Neumann proof, if you actually come to grips with it falls apart in your hands! There is nothing to it. It's not just flawed, it's silly! [...] The proof of von Neumann is not merely false but foolish!''\footnote{As cited in \citep[88]{mermin1993hidden}.}. Thirty years later \citet{mermin1993hidden}, following Bell, still considers that  ``von Neumann's no-hidden variables proof was based on an assumption that can only be described as silly''. Going  forward in time another 17 years,  Jeff \citet[1334]{bub2010neumann} argues that ``Bell's analysis misconstrues the nature of von Neumann's claim, and that von Neumann's argument actually establishes something important about hidden variables and quantum mechanics''.

The details of the von Neumann no-go theorem will not concern us here, but this example of a history of a single no-go theorem nicely illustrates the difficulty of interpreting no-go results in physics. Opinions about it varied between having established a ``decisive answer'' on the question of hidden variables  to the proof being considered  ``foolish''; the whole debate now ranging more than eight decades. This is not to say that there was no progress or that there is not a way to understand the disagreement and its development. However, this example  illustrates that the role of no-go theorems in physics  seems to differ from the case of impossibility results in mathematics. When we prove something in mathematics, there usually does not seem to be that much disagreement about what the theorem implies. This already hints at the more complex structure of  no-go theorems in physics compared to those in mathematics. There is a plethora of examples in the  history of physics where this more complex structure was not adequately recognised and where it was misunderstood what no-go theorems can imply. In this paper we want to analyse abstractly the general  implications one can draw from no-go theorems and the role they can serve in theory development.

 It is not in terms of the content of the theorems that the analysis proceeds --  although it will have significant implications for it --  it is in terms of an understanding of no-go theorems as methodological tools in theory development. As such, it is crucial to understand what capabilities no-go theorems can have. As being unaware of what the general role of no-go theorems can be, i.e. of what they can possibly imply, bears the danger of misinterpreting what a particular result actually implies and can misdirect a whole research effort based on a misinterpretation of the situation. The aim of this paper thus  is to address this more general argument structure of no-go theorems. 

\comment{
Talking about  no-go theorems in all generality bears a danger. Are all no-go theorems similar enough to analyze them in all generality? 
However,
}

 I start in Sect. \ref{case}  with the presentation of a case study of a set of no-go theorems from particle physics, which serves as an illustration of the various elements of a no-go theorem and subsequently allows us to provide an analysis of the abstract argument structure of no-go theorems (Sect. \ref{structure}). In Sect. \ref{imply},  \ I discuss the methodological consequences of a no-go theorem for each individual element in more detail. In Sect. \ref{whereto}, I consider what the previous analysis implies for no-go theorems more generally and how one should adequately interpret the result of a no-go theorem.




%
%

\section{The Development of a No-Go Theorem: Combining Internal and External Symmetries}\label{case}

Our tactic in assessing what no-go theorems imply, is to start by considering a specific historically rich development of a set of no-go theorems from which we can establish the various components relevant for the more abstract discussion. More specifically, it is a not much discussed example of a set of no-go theorems from particle physics, each aiming to establish the impossibility to combine internal and external symmetries.


Symmetry transformations can act on  different degrees of freedom of the physical system. External symmetries refer to those symmetries that act on the spatiotemporal degrees of freedom.  These can be the discrete symmetries of parity and time reversal or continuous symmetries like translations and boosts. The spacetime symmetry that physicists  were concerned with in the examples we will consider, focused on the Poincar\'{e} symmetry, which contains the Lorentz symmetry and the symmetry under translation. One contrasts external symmetries with internal symmetries. Internal symmetries are symmetry transformations that do not act on the spacetime degrees of freedom but rather on an ``internal'' space.  Examples are  Gell-Mann and Ne'eman's $SU(3)$-flavour symmetry, which in modern terms, is a symmetry under the change of the flavour of quarks with respect to the strong force. Other popular examples are Heisenberg's $SU(2)$-Isospin of the neutron and proton or the standard model gauge group $SU(3) \times SU(2) \times U(1)$.\footnote{To schematically illustrate it: in a field theory, external symmetries refer to transformations of the form $\Phi_I(x) \to \tilde{\Phi}_I(\tilde{x})$ under $x \to \tilde{x}$ and internal symmetries to transformations of the form $\Phi_I(x) \to \tilde{\Phi}_I(x)$.}  \\
I will discuss two motivations for why physicists tried to combine internal and external symmetries (\ref{why}). This will be followed by a discussion of some no-go theorems that culminated in the result of Coleman and Mandula in 1967 (\ref{ng}). Finally, I will discuss certain routes towards combining internal and external symmetries which were not affected by the no-go theorems (\ref{hls}).\footnote{See \citep[Sect. 24]{weinberg2011supersymmetry} and \citep{di2000notes} for  historical accounts and  \citep{iorio5221alternative} for a more systematic treatment, parts of which we follow here.}


\subsection{Why Combine Internal and External Symmetries?}\label{why}

Symmetries in physics are strongly related to the properties characterising the particles of the theory.
To put it  briefly: one looks for those operators that commute with the generators of the symmetry. The eigenvalues of these operators then correspond to the invariant properties of the particles.\footnote{When we speak of ``particles'', it should not be interpreted as a statement about our commitment with regard to the ontology of quantum field theory. It is used here in the usual particle physics parlance (see also the physics literature cited below). More precisely, the eigenvalues of the Casimir operators, i.e. the commuting operators, are the invariant properties, understood as uniquely determining the irreducible representations of a group. These irreducible representations are associated with what we call ``particles'' above. My thanks to an anonymous referee for pointing me to this possible point of confusion.}  The properties thus related to the Poincar\'{e} group, i.e. the external symmetry, are spin and mass. For internal symmetries like SU(2) it is the isospin or for SU(3) it is the quark flavour. 
One can always combine internal and external symmetries trivially, by considering the direct product of the two groups. In this case, however, all elements of the internal and external group commute with each other and so remain independent. One is therefore interested in the non-trivial combinations of the symmetry group, i.e. a group which combines the operators of the symmetry groups in such a way that they do not commute. There were two main motivations behind the wish to combine internal and external symmetries, which we now turn our attention to.

\paragraph{The Problem of Mass-splitting} The first motivation to combine internal and external symmetries was to account for the mass gap between protons and neutrons.  \citet{heisenberg1932iuber} introduced the $SU(2)$-Isospin symmetry between protons and neutrons to account for their equal interaction under the strong force\footnote{Although they do not interact equally under electromagnetic interactions, as the neutron is neutrally charged  and the proton positively charged.}. This internal symmetry transforms between protons $|+ \rangle$ and neutrons $|-\rangle$, i.e. $I_\mp | \pm\rangle = | \mp\rangle$ with $[I_+,I_-]=2I_0$ and $I_0 | \pm\rangle = \pm | \pm\rangle$.   

\comment{
The corresponding algebra can be written in terms of operators satisfying:
\begin{equation*}
[I_+,I_-]=2I_0 \qquad   [I_0,I_{\pm}]=\pm I_{\pm}
\end{equation*}
with 
\begin{equation*}
I_0 | \pm\rangle = \pm | \pm\rangle, \qquad  I_\pm | \pm\rangle = 0, \qquad I_\mp | \pm\rangle = | \mp\rangle.
\end{equation*}
}
The translation generator of the Poincar\'{e} group $P_\mu$, an external symmetry generator, commutes with the $I_{\pm}$, i.e. $[P_\mu,I_\pm]=0$. From this it follows that 
$P^{\mu}P_{\mu} |\pm\rangle=m^2|\pm\rangle$, where $m$ is the mass of the states. That is, since the momentum generator commutes with the $SU(2)$ generator, the proton and neutron will have to have the same mass. Although this is a good approximation, protons and neutrons do not have the same mass. The idea was then that a non-trivial commutation relation between them may lead to the known mass difference between the proton and the neutron. For instance, by assuming $[P_\mu,I_+]=c_\mu I_+$ one obtains after some manipulations using the changed commutation relations $P^2 |+\rangle = I_+ P^2|- \rangle - c^2 |+\rangle$ from which one can easily show  $m_p^2=m_n^2-c^2$. One can then recover the hoped for mass difference by experimentally fixing the $c^2$ value. So by mixing internal and external symmetries the hope was to explain the mass difference of particles. This initial motivation turned out not to be significant, as nowadays we know that protons and neutrons are composite particles made up of different quarks.


\paragraph{Unification}
The second motivation for combining internal and external symmetries is the  methodological urge within the particle physics community to unify. If  internal and external symmetries could be understood as following from one more general unified simple group, we would be one step further in the unification program within particle physics.
 Consider \citet{gell1964schematic} and \citet{ne1961derivation}'s $SU(3)$-Flavour Symmetry. During the 1960s many new particles were being discovered and the relation between them was unknown. It was the SU(3)-flavour symmetry that allowed an understanding of the different baryons and mesons then discovered as elements within multiplets of the same  group. There is for instance a baryon octet that combines particles with different strangeness and charge but the same spin, namely spin-$\frac{1}{2}$. Similarly, there is a baryon decuplet  combining spin-$\frac{3}{2}$ particles  into one multiplet. \\ 
Having unified particles with different strangeness and charge within multiplets the hope was to be able to unify particles with different spins within one multiplet as well. Since spin is a property related to an external symmetry, this would amount to combining internal (strangeness, charge) and external (spin) properties. So bringing particles with different charges, strangeness and spins within a multiplet can be achieved by bringing together internal and external degrees of freedom in a non-trivial way.
One early step in this direction  was the SU(6) symmetry group. The $SU(6)$ group was introduced and succeeded  in unifying the baryon octet and decuplet into a 56-plet\footnote{See e.g. \citep{sakita1964supermultiplets} and \citep{gursey1964spin}.}. This gave rise to further attempts at unifying internal and external symmetries, since $SU(6)$ was not yet the end of the story. What $SU(6)$ achieved was a unification of $SU(3)$-flavour with non-relativistic $SU(2)$ spin. A full relativistic unification, i.e. one including the full Poincar\'{e} group, was then hoped for and attempted. But attempts failed, leading the way to several no-go theorems.

\subsection{No-Go Theorems} \label{ng}
Several no-go theorems were proposed between 1964 and 1967 culminating in the famous Coleman-Mandula theorem. The no-go theorems that were being developed ranged from  mathematical to more and more physical arguments for the impossibility of combining internal and external symmetries. We will now mention three no-go theorems starting with the simplest argument made by \citet{McGlinn:1964ji} for the impossibility of combining internal and external symmetries.\\
In 1964 McGlinn, having the mass splitting problem from before in mind,   proved the following theorem\footnote{We follow O'Raifeartaigh's presentation of McGlinn's theorem in \citep{o1965lorentz} to allow for a more coherent nomenclature.}.
\begin{quote}
\textbf{McGlinn Theorem:} Let $\mathcal{L}$ be the Lie algebra of the Poincar\'{e} group, $M$ and $P$ the homogeneous and translation parts of $\mathcal{L}$, respectively, and $\mathcal{I}$ any semisimple internal symmetry algebra. 
\begin{description}
\item[(a)] If  $\mathcal{T}$ is a Lie algebra whose basis consists of the basis of $\mathcal{L}$ and the basis of $\mathcal{I}$, and
\item[(b)] if $[\mathcal{I},M] = 0$ (i.e. the internal symmetry is Lorentz invariant) 
\end{description}
then $[\mathcal{I} , P ] = 0$. Hence $\mathcal{T} = \mathcal{L} \times \mathcal{I}$. 
\end{quote}
So if (a) and (b) are satisfied, one can combine the internal group  $\mathcal{I}$ with the external group $\mathcal{L}$ only trivially. Note this is a mathematical result, in the sense that it is not a result that follows from within the framework of a physical theory. As such it seems to be of a more general nature.

McGlinn's theorem gave rise to several papers which aimed to weaken the assumptions. For instance, early attempts by \citet{michel1965relations} and \citet{sudarshan1965concerning} showed that to obtain McGlinn's result, it is sufficient to assume that only one of the generators of the internal symmetry algebra $\mathcal{I}$ does not commute in (b). But it is especially assumption (a) that seems too stringent and unnecessary and which therefore motivated O'Raifeartaigh in 1965 to prove  a more general theorem. Rather than building up the larger group starting from the Poincar\'{e} group, O'Raifeartaigh looked for the most general way to embed the Poincar\'{e} group into a larger group, with the only restriction that the larger group is of finite order. The finite order of the larger group is necessary so that the so-called Levi decomposition theorem, which forms the basis of his theorem, can be applied.  So with the only requirement that the group within which the Poincar\'{e} group is to be embedded  be of finite order, O'Raifeartaigh was able to categorise the possible embeddings in the following theorem:
\begin{quote}
\textbf{O'Raifeartaigh Theorem:} 
Let $\mathcal{L}$ be the Lie algebra of the Poincar\'{e} group, consisting of the homogeneous part $M$ and the translation part $P$. Let $\mathcal{T}$  be any Lie algebra of finite order, with radical  $S$ and Levi factor  $G$. If $\mathcal{L}$ is a subalgebra of $\mathcal{T}$, then only the following four cases occur: 
\begin{enumerate}
\item[(1)] $S = P$;
\item[(2)]  $S$ Abelian, and contains $P$;
\item[(3)] $S$ non-Abelian, and contains $P$;
\item[(4)] $S \cap P = \emptyset$. 
\end{enumerate}
In all cases, $M \cap S = 0$.\footnote{Some background may be helpful here: the Levi decomposition theorem states that any Lie algebra of finite order can be decomposed into the semi-direct sum of its radical (maximally solvable Lie algebra) and Levi factor (semisimple Lie algebra). Since $P$ is abelian its first-derived algebra is empty and therefore solvable. $M$ is semisimple therefore not solvable and contained in $G$. This leads to the four mentioned possible cases of decomposition.}
\end{quote}
O'Raifeartaigh then goes on to discuss each possibility in detail. One thing that one can already see is that from a purely mathematical point of view, it is possible for the internal  and external symmetry to be combined in a non-trivial way. O'Raifeartaigh shows that case (1) reduces to the McGlinn case of a trivial combination, where one obtains $\mathcal{T} = \mathcal{L} \times \mathcal{I}$. In the other three cases (2)-(4), the internal and external symmetries could possibly be combined non-trivially but are, as O'Raifeartaigh argues, physically unreasonable. For instance, case (2) necessitates a translational algebra of more than four dimensions, or case (3) has the problem that, due to Lie's theorem, any finite dimensional representation of  a solvable non-abelian algebra has a basis such that all matrices have only zeros above the diagonal, i.e. are triangular matrices. This leads to the problem that one cannot always define hermitian conjugation. So unlike McGlinn's theorem, O'Raifeartaigh's theorem rules out a non-trivial combination of internal and external symmetries for physical reasons.\\
Although O'Raifeartaigh was able to generalise McGlinn's no-go theorem it was still considered to have shortcomings. One shortcoming was the need to consider only Lie algebras of finite order and  the second shortcoming is the  concentration on only the one-particle spectrum. \citet{coleman1967all} were able to account for both of these shortcomings by moving away from the mathematical framework of McGlinn and O'Raifeartaigh, towards a physical framework, namely S-Matrix theory, wherein the symmetries from before  are the symmetries of the S-matrix.\footnote{Coleman was already working on the problem of combining internal and external symmetries in 1965 when he was able to show that certain relativistic versions of $SU(6)$ had absurd consequences and should therefore be discarded \citep{coleman1965trouble}.} This allowed them to consider n-particle spectra but still without the need to consider any specific quantum field theory. Also no need for finite order Lie algebras was necessary anymore.  However, several physical and mathematical assumptions were introduced. The Coleman-Mandula Theorem states the following:
\begin{quote}
\textbf{Coleman-Mandula Theorem:} Let $\mathcal{T}$ be a connected symmetry group of the $S$ matrix, and let the following five conditions hold: 
\begin{enumerate}
\item $\mathcal{T}$ contains a subgroup locally isomorphic to the  Poincar\'{e} group $\mathcal{L}$; 
\item all particle types correspond to positive-energy representations of $\mathcal{L}$, and, for any finite mass $M$, there are only a finite number of particle types with mass less than $M$;
\item elastic-scattering amplitudes are analytic functions of the center of mass energy and of the momentum transfer in some neighbourhood of the physical region;
\item at almost all energies, any two plane waves scatter; 
\item the generators of $\mathcal{T}$ are representable as integral operators in momentum space, with distributions for their kernels.
\end{enumerate}      
Then $\mathcal{T}$ is locally isomorphic to $\mathcal{L} \times \mathcal{I}$ , the direct product of the Poincar\'{e} group and the internal symmetry group.
\end{quote}
This represented the final blow to attempts in the community at unifying internal and external symmetries.\footnote{With a single exception: \citet{mirman1969physical} made the more general claim that ``the impossibility theorems have no physical relevance''. This was followed by \citet{cornwell1971relevance}, where it is claimed that ``Mirman's objections may be overcome without difficulty, and that the above-mentioned theorems do indeed relate to the physical situation''.}  It is interesting to note that the physicists working on this unification project were actually hoping for the opposite result. While aiming for unification they apparently ended up showing its impossibility.


\subsection{The Rise of Supersymmetry} \label{hls}
As mentioned, the Coleman-Mandula theorem stopped much of the discussion on internal and external symmetries. The explicit assumptions above did not give rise to physicists attempting to weaken the assumptions, although some problems with them were known (see e.g. \cite{sohnius1985introducing}). 
However, in the subsequent years, three  different groups with completely different motivations were able to non-trivially combine internal and external symmetries.
The first successful proposal was by Yuri Golfand and his student Evgeni Likhtman from the Physical Institute in Moscow.\footnote{See \citet{Golfand:1971iw} for the original paper and  \citet{golfand1972extensions} for an elaboration on the 1971 paper.} The actual reason motivating Golfand to develop an extension of the Poincar\'{e} group is not clear.  However, they try to account for parity violation in the weak interactions in their original paper. Although,  they also state the following reason: "only a fraction of the interactions satisfying this requirement [i.e. being invariant under Poincar\'{e} transformations] is realised in nature. It is possible that these interactions, unlike others, have a higher degree of symmetry" \citep[p.323]{Golfand:1971iw}. So the search for this higher symmetry can be seen to have been their goal as well. \citet{Volkov:1972jx} from the Kharkov Institute of Physics and Technology had other reasons for their work. They hoped to be able to describe the neutrino, then thought to be massless, as a Goldstone particle. Obtaining Goldstone particles with half-integer spin like the neutrino makes an extension of the Poincar\'{e} group with spinorial generators necessary. And finally, \citet{wess1974lagrangian} discovered a 4D supersymmetric field theory by trying to extend the 2D version obtained in String Theory. The results were not affected by the Coleman-Mandula result. In fact, none of the papers even referred  to the Coleman-Mandula theorem, since none of them were motivated by the aim to combine internal and external symmetries.\footnote{Only in a second paper, did Wess and Zumino note in a footnote that  ``[t]he model described in this note, and in general the existence of supergauge invariant field theories with interaction, seems to violate $SU(6)$ no-go theorems like that proven by S. Coleman and J. Mandula [...]. Apparently supergauge transformations evade such no-go theorems because their algebra is not an ordinary Lie algebra, but has anti-commuting as well as commuting parameters. The presence of the spinor fields in the multiplet seems therefore essential''  \citep{wess1974supergauge}.}

So how did they do it? An implicit assumption of the Coleman-Mandula no-go theorem is the use of Lie algebras to represent the symmetries, a mathematical assumption, which turned out to be too restrictive. Golfand and Likhtman, Akulov and Volkov as well as Wess and Zumino introduced, without explicitly realising it, a more general mathematical structure to represent symmetries, so called graded Lie algebras. A structure which was introduced in the mathematics literature in the mid-1950s\footnote{The first paper introducing it was \citet{nijenhuis1955jacobi}. See \citet{corwin1975graded} for an excellent review article on the application of graded Lie algebras in mathematics and physics.}. This more general mathematical structure allowed them to non-trivially combine internal and external symmetries in what is nowadays called supersymmetries.

\section{Modelling No-Go Theorems Abstractly}\label{structure}


In Sect. \ref{case} we have  seen  the history of a set of no-go theorems, from early motivations to how it was finally circumvented. It was chosen as a case study, as it provides us with enough detail to model no-go theorems more abstractly and identify the relevant elements involved in their assessment.



No-go theorems usually start with a goal $G$. 
One e.g. aims to unify internal and external symmetries,  find a hidden variable theory or  simulate neutrinos. The no-go theorem then purports to show that achieving this goal is not possible. Once the goal is determined the no-go theorem is set within a certain framework $F$, which is usually chosen based on its suitability to achieve $G$. So for some purpose one may not need to consider a specific theory within which one tries to show the impossibility of G but may wish to do so on purely mathematical grounds from which one then infers that it generally holds. Thus, the framework can be a mathematics-framework (as in the McGlinn and O'Raifeartaigh no-go theorems), a theory-framework (like the use of S-matrix theory by Coleman and Mandula), or a model-framework (based  e.g. on toy models or possible extensions of existing theories, as in the derivation of the Bell inequalities). Within the framework one is then able to phrase the physical assumptions $P$ that are represented by certain mathematical  structures $M$. $M$ for our purposes will contain both the mathematical structures used to represent the physical assumptions (e.g. Lie groups, Kolmogorovian probabilities, etc.) as well as the mathematical tools and methods used to derive the result.\footnote{The elements $F$, $P$ and $M$ are recognized in formally more careful reconstructions of quantum theory. See for instance \cite{hardy2001quantum} and \cite{clifton2003characterizing} for reconstructions and \cite{grinbaum2007reconstruction} for a philosophical discussion. I thank an anonymous referee for pointing me to these.} 

In a no-go theorem one derives from $F$, $P$ and $M$ something which either contradicts $G$ directly or establishes $G$ by violating another physical background assumption $B$. Taking $B$ into account is important as we saw in O'Raifeartaigh's theorem. There, one is actually able to combine internal and external symmetries but will then  not be able to define hermitian conjugate operators, which are needed to guarantee real eigenvalues that correspond to physical quantities in quantum mechanics. Similarly, in the case of Bell's no-go theorem, one considers the consequences of a generic hidden variable theory, which lead to the Bell inequalities, and how they disagree with the confirmed predictions of quantum mechanics. So the goal $G$ of obtaining a hidden variable theory has been satisfied, while it disagrees with the physical background assumptions $B$, i.e. the predictions of quantum mechanics, which were not part of the derivation of the inequality. We have now all the components necessary to give an abstract definition:

\begin{quote}
Definition: A \textit{No-go result} has been established iff an inconsistency  arises between 
\begin{itemize}
\item a derived consequence of a set of physical assumptions $P$ represented by a mathematical structure $M$ within a framework  $F$, 
\item and a goal $G$ or a set of physical background assumptions $B$.
\end{itemize}
We denote an abstract no-go result with $\langle  P, M, F \lightning G, B \rangle$.
\end{quote}

The arrow, $\lightning$, denotes the contradiction between $P, M, F$ on the one side and $G$ and possibly $B$ on the other.  
The physical assumptions and the mathematical structures used to represent them are, of course, strongly dependent on each other. Obviously all elements $G$, $B$, $F$, $P$ and $M$ are dependent on each other to some extent and one may argue that it seems not obvious how to demarcate, for instance, $P$ and $M$. But as we will see -- and as our aim is to follow a methodological goal --  it is reasonable to  distinguish between them, since in most cases one can change the individual elements separately. For example I can go from a  mathematics-framework to a theory-framework while still considering the physical assumption of using certain spacetime symmetries and using for that purpose the mathematical structure of Lie algebras. However, as we saw in the case of the Coleman-Mandula theorem, going from one framework to the other (from a mathematics-framework to a theory-framework) still made it necessary to add additional assumptions to establish the no-go result. This exemplifies that one may separately change the assumptions involved; however, these changes will usually not be independent from changes in the other assumptions.

The historical case study  does not force the above definition upon us. The justification for defining no-go theorems in the above sense, and to model its elements as above, comes from its methodological fruitfulness and the wish to stay close to scientific practice, which in turn brings in some vagueness in the individual elements and their logical and structural relations. For many no-go theorems, however, the above definition is readily and fruitfully  applicable as will be illustrated in what follows.  It is not the aim of the paper to establish that the above definition is applicable to all no-go theorems (an impossible task). The elements defined above are quite broad in their intended domain and so may encompass more than they bargained for: theorems that are  usually not considered no-go theorems may also fall under the above definition.  This would not weaken in any way the methodological implications I want to draw from no-go theorems, but only weaken the use of the above definition to pick out no-go theorems among all theorems. A task that I do not aim to address in the paper, as a no-go theorem is a specific kind of theorem in physics that is  distinguished from other theorems not structurally but in their purpose. They purport to establish the impossibility of something and the above explication serves to account for this purpose. The aim of this paper is to establish under what conditions no-go theorems  can, if at all, serve this purpose. 

\section{The Different Elements of No-Go Theorems} \label{imply}

In this section we want to discuss each element of $\langle  P, M, F \lightning G, B \rangle$ in more detail.
No-go theorems construed as above are contradictions. So to resolve the contradiction one has to deny at least one of its elements. These denials amount to a methodological step in the use of no-go theorems in theory development.
 For example, some no-go theorems have  had the impact of stopping whole research programs. In these circumstances they were understood as showing the impossibility of $G$ only. In other circumstances they made  certain assumptions explicit and showed through that a methodological pathway in how to go about achieving G, by denying one of the other assumptions.
Given the   structure we have established, it is legitimate to assess the viability of denying each element and what methodological pathway that amounts to. For that purpose we need to  consider the different elements more closely, analyse the possible justifications we may have for each and consider the possible implications we may draw from their denial. 
We use the following notation:
\begin{equation*}
\langle  P, M, F \lightning G, B \rangle \Rightarrow  \neg G \lor  \neg P \lor  \neg B \lor  \neg F \lor  \neg M.
\end{equation*}
While we do use the logical notation, i.e. $\neg$ and $\lor$, one should  understand the above symbolically, pointing to different possible \emph{methodological pathways} rather than strict logical implications, pointing to a strict independent denial of either one of the disjuncts. 


\subsection{Methodological Pathway 1: $\langle  P, M, F \lightning G, B \rangle \Rightarrow \neg G$: }

Here  the no-go result is interpreted as the impossibility of $G$. This is for example  how von Neumann's no-go theorem was understood for thirty years or the Coleman-Mandula theorem till the advent of supersymmetry. Both stopped whole research programs. Although, given the general structure of no-go theorems, concentrating on the denial of $G$ may seem odd, but it is not too surprising. $G$ is some goal, which obviously is not yet established, while the other elements are at least perceived to be part and parcel of the well-confirmed physics. But if $G$ is not  part and parcel of the well-confirmed physics, why is it considered to be a goal in the first place? This, of course, leads us to the issue underlying motivations behind theory development.\footnote{Of particular interest, since they are formulated in a language close to scientific practice, is  \cite{laudan1978progress} and \cite{nickles1981problem}. } For our purpose we will consider the following list of possible motivations for setting goals for theory development.


\underline{Empirical Motivation:}
One motivation for setting a goal $G$ might be some empirical observation,  which existing theories cannot adequately accommodate.
We saw that one motivation for combining internal and external symmetries was the unexplained observed mass difference between the proton and the neutron. Combining internal and external symmetries was a possible way to address this. \\
\underline{Metaphysical Motivation:}
A goal may be motivated by metaphysical considerations. One way of understanding the program of completing quantum mechanics, i.e. to provide  a hidden variable theory, is metaphysical. Finding a theory of hidden variables  is not necessitated by some observed phenomenon that quantum mechanics cannot account for. One may argue that  it is motivated by the hope to find an ontologically coherent understanding of its domain of applicability.
\\
\underline{Meta-inductive Motivation:}
The second motivation we discussed as to why to combine internal and external symmetries was unification (combining particles of different spin into one multiplet). Unification is also not necessitated by some empirical observations, but is often considered  a successful ingredient in theory development. One may argue, see e.g. \cite{maudlin1996unification}, that unification is meta-inductively motivated, i.e. one infers from previous successes of attempts at unification to future ones. 
\\

\underline{Pragmatic Motivation:}
Another possible motivation can be purely pragmatic. Consider for instance the theorem that \citet{nielsen1981absence} proved. They show that neutrinos, or more generally chiral fermions, cannot be simulated on a lattice. So this result puts certain \emph{calculational} limitations on simulating certain phenomenon in particle physics.  As the aim of lattice gauge theories are to do certain calculations, which are otherwise very difficult, there is nothing of great foundational significance about this theorem. The original goal was pragmatically motivated.
\\[10pt]

These are possible motivations one may give for some goal $G$. There is no claim regarding the completeness of this list. The relevant point is that there may be different motivations for $G$ and different motivations may lead to different implications one may want to draw from the no-go result. Note that there are cases where one and the same $G$ is motivated by different theorists for different reasons, c.f. the two motivations from \ref{why}. Accordingly, the implications of the same no-go theorem may differ for these different theorists. For instance, it seems obvious that a goal which is metaphysically motivated may lead to a different interpretation of the theorem compared to one that was  motivated purely pragmatically.  \citet{laudisa2014against}, for instance, argues against the significance of many recent no-go theorems in quantum mechanics. He claims that the ``search for negative results [...] seems to hide the implicit tendency to avoid or postpone the really hard job'', which for him is partly ``to specify the ontology that quantum theory is supposed to be about'' \citep[p.16]{laudisa2014against}. However, one can understand the programme of finding a hidden variable theory, both as a metaphysical programme as well as a programme of finding a probabilistic foundation of quantum mechanics\footnote{For example, Arthur \citet{fine1982joint} followed this second route with generalised probability spaces. See also more recent  discussions in   \citep{suppes1991existence,hartmann2014imprecise,feintzeig2017noncontextual}.}. The significance of one and the same no-go theorem, like e.g. the Bell inequality, will therefore be differently assessed depending on one's motivations for that goal G.

Besides different motivations, also being insufficiently explicit about the goal $G$ can lead to confusion about the evaluation of the no-go theorem.  Note that  $\langle  P, M, F \lightning G, B \rangle$ does not imply $\langle  P, M, F \lightning G', B \rangle$ when G implies G$'$. This is nicely illustrated by two  recent papers by \citet{cuffaro2017significance,cuffaro2018reconsidering}. While discussing the Bell inequality and the GHZ equality, he distinguishes between two kinds of context: the theoretical and the practical context. Within the theoretical context one may consider the Bell result to shed light on the questions of whether there is an alternative locally causal theory of the world able to replace quantum mechanics. In the practical context, one may ask whether one can  classically reproduce, by e.g. a classical computer simulation, the predictions of quantum mechanics. These two contexts are very different. As Cuffaro shows, a denial of the goal in the theoretical context does not imply a denial of the classical simulability of the considered quantum correlations. The reason why we would still reject those in the other context is because a ``set of plausibility constraints on locally causal descriptions [...] in the context of this question is implicitly understood by all'' \cite[p. 634]{cuffaro2018reconsidering}. 

Finally, it is important to note that the goal $G$ does not need to always be desirable. In fact, there are cases where the no-go theorem is established  to rule out the possibility of $G$. Coleman and Mandula  did not have the desire to show that you can combine internal and external symmetries, but they wanted to conclusively show its impossibility (even though, as we saw, they failed to do so).




\subsection{Methodological Pathway 2: $\langle  P, M, F \lightning G, B \rangle \Rightarrow \neg P \lor \neg B$: }
Let us turn to the physical assumptions. I include the physical background assumptions $B$ here as well as they are after all physical assumptions. However, unlike $P$, if they are included at all, it is usually as a crucial assumption that is much more supported. So for that purpose we will not consider them explicitly in what follows. 
A no-go result that is not understood as having established the impossibility of $G$, is quite commonly understood as an impossibility result with respect to the physical assumptions $P$. It is  usually with respect to one single assumption $p \in P$, if one considers that one assumption to be the least defensible. This is the situation when Einstein, Podolsky and Rosen (\citeyear{einstein1935can}) infer the incompleteness of quantum mechanics rather than denying the physical assumption of locality.

Physical assumptions need to be discussed case by case and a general discussion will not allow us to draw concrete conclusions, but we can still recognise that there are physical assumptions of different kind. Obviously, the goal $G$ determines to a large extent the physical assumptions. If my goal is to combine the Poincar\'{e} group with some internal group, then trivially I will take as one of my physical assumptions that one of the groups adequately represents the assumed symmetries of space and time. 

There are also physical assumptions that are part of  well-confirmed theories, like energy conservation, or physical assumptions that have been introduced for the sole purpose of deriving the result. An example is the analyticity assumption of Coleman and Mandula (assumption 3 above). 

As one can see, these different physical assumptions are not comparable in terms of the justification one can give for them. While some assumptions can be justified empirically, others cannot, and may correspond to metaphysical positions\footnote{The Reality criterion of \citet{einstein1935can} may be read as such. However, its status is still debated: see \cite{Maudlin_2014,Werner_2014,glick2019reality}.} and external requirements on what the future theory needs to satisfy. So while we may say that we have evidence supporting the claim that energy is conserved, we may not want to claim the same for the reality criterion in the Einstein-Podolski-Rosen  setup or the factorisability assumption in the Bell inequalities. These are cases where much disagreement about the possible importance and justification for the assumption can arise and where most of the philosophical debate of no-go theorems is understandably situated. This is important, as careful analysis of these assumptions are sometimes lacking in the physics literature. For instance,  \citet[p.159]{coleman1967all} claim that the analyticity assumption ``is something that most physicists believe to be a property of the real world''. This, one may reasonably argue, needs further discussion.

One strategic option used in the context of physical assumptions is to replace one physical assumption by  a weaker physical assumption. If I consider for instance some $P_1$ to be the least defensible of the assumptions, I may give up less by further distinguishing that assumption by its possible conjuncts. That is, if I follow the route of $\neg P_1$ I may consider that to entail $\neg P_1= \neg (P_{1a} \land P_{1b})= \neg P_{1a} \lor \neg P_{1b}$. So it would suffice to give up $P_{1a}$ or  $P_{1b}$  and thereby giving up something weaker. This, however, does not entail that these weaker assumptions are then safe from other possible no-go theorems, but only that that specific no-go result is affecting it. An example of this strategy in play is the consideration of  the factorisability assumption of the Bell inequalities as a conjunct of the assumptions of parameter independence and outcome independence as introduced by \citet{jarrett1984physical}.




%
%
%

\subsection{Methodological Pathway 3: $\langle  P, M, F \lightning G, B \rangle \Rightarrow \neg F$: }

No-go theorems in physics are not always formulated within  a theory (e.g. the standard model of particle physics or thermodynamics).  As we saw in the examples from the last section, the McGlinn theorem as well as the O'Raifeartaigh theorem are theorems, which are theory independent, as they can be seen as results of group theory. The possibility to frame a no-go theorem in physics outside of specific theories points to an additional element I would like to make explicit, namley the framework $F$. The McGlinn and O'Raifeartaigh theorem are within a  mathematics-framework. That is, one considered two mathematical structures and asked whether there is a mathematical structure that non-trivially combines them. On the other hand, Coleman and Mandula's theorem is a result within a theory, namely S-matrix theory. They were considering the external and internal symmetries as symmetries of the S-matrix and so chose a theory-framework for their no-go result. In other cases, one may develop a model and prove within that model-framework the no-go result.  

The framework $F$ of a no-go result has not played much of a role in the evaluation of no-go theorems.  This can be due to the apparent neutrality of the framework with respect to the no-go result. In most cases it seems that the choice of framework is  fixed by the  \emph{kind} of goal one aims to reach rather than the specific goal itself. If I aim  to find a hidden variable account of quantum mechanics, I start by building a general model on which I impose the physical properties (elements of $P$) of the desired hidden variable account. So I choose a model-framework, which may still lack the details of the dynamics of the theory etc. It is, at least at first, not clear how a theory-framework or a mathematics-framework could be helpful here.
Similarly, it seems to be a mathematical issue, whether one can combine two symmetry groups non-trivially.  So combining them without any specific theory in mind seems to be the obvious and more general approach. So one chooses a mathematics-framework. The move towards S-matrix theory, i.e. a theory-framework, was not based on not being satisfied by the mathematics-framework but was largely motivated by the aim to weaken the strong assumption of restricting oneself to finite parameter groups made by O'Raifeartaigh and it was  not  obvious how the theorem could have been extended to infinite parameter groups as it relied so strongly on Levi decomposition.   

The above example nicely illustrates that the framework is mainly chosen for pragmatic reasons and is not independently justified. However, using different frameworks may still provide us with different perspectives. \citet{pitowsky1989quantum}, for instance, provides a different perspective on the hidden variable program and the Bell inequalities\footnote{A result derived within a model-framework.}. He shows that one can understand the question whether a set of probabilities are classical (Kolmogorovian) probabilities, not only by considering whether they satisfy the Kolmogorovian axioms, but also, equivalently, by whether they satisfy a set of inequalities. He shows that the inequalities for certain classical setups correspond to Bell-type inequalities.  Inserting probabilities predicted by quantum mechanics for certain quantum mechanical experiments\footnote{Note that not all quantum mechanical experiments lead to a violation of the Bell inequalities.} into the inequality leads to a violation of that inequality. However, unlike the implications in the model-framework, one infers in the mathematics-framework to the comparably more mathematical conclusion that not all quantum mechanical experiments have classical probability space representations.



A reason why the significance of the framework $F$ has not been important in the evaluation of no-go results is the lack of an obvious interpretation for $\neg F$. In the case of the goal $G$ and the physical assumptions $P$, the denial could be understood as their respective impossibility. This is usually not so for the framework. It does not make sense to talk of the impossibility of a certain mathematics-framework or  model-framework, but only of the assumptions realised within it. 
However, an understanding of the respective negations as opening up possible methodological pathways provides important strategic options. The benefit of considering a change of framework has already been illustrated in the case of the hidden variable program, where the move to a mathematics-framework presented a new perspective. The new perspective, however, came effectively with a different goal, more concerned with the probabilistic foundations rather than a locally causal hidden variable theory. These are strongly dependent questions, however, with different foci and thereby opening up different methodological pathways.

Finally, there are not only options of going from one kind of framework to another but also options within one kind of framework. A result obtained within one theory may or may not hold for another theory. This is even the case with different formulations of the same theory. We can consider a no-go result we obtain in one formulation to also hold in the other, only if we have reason to believe that they are equivalent in the relevant sense. However, the Coleman-Mandula result is a result within S-matrix theory, and it is, for example, not obvious that it will similarly hold within Lagrangian quantum field theory, as the symmetries of the S-matrix are not necessarily symmetries of the Lagrangian. Similarly for results within classical mechanics, the differences in the Lagrangian and Hamiltonian formulations have been a much discussed topic in philosophy of physics\footnote{See for instance \citet{north2009structure}, \citet{curiel2014classical} and \citet{barrett2015structure}.}.

\subsection{Methodological Pathway 4: $\langle  P, M, F \lightning G, B \rangle \Rightarrow \neg M$: }

Let us turn to the last crucial element of no-go results, the mathematical structure $M$, which encompasses the mathematical structures, tools and methods as well as the underlying logic.\footnote{The underlying logic has played an important role in discussions surrounding the  logic of quantum mechanics. I thank Hartry Field for pointing me to this.} There are usually many necessary mathematical assumptions involved in the derivation of a no-go result. For example, assumption 5 of the Coleman-Mandula theorem is of this kind. It is an assumption that Coleman and Mandula admit is ``both  technical and ugly'', and for which they hope ``that more competent analysts will be able to weaken [...] further, and perhaps even eliminate [...] altogether'' [p.159]. There may also be additional assumptions involved in the  derivational steps, like the use of certain approximation methods and limits. All of these can possibly be problematic and should be carefully assessed.  However, we will focus on another element of $M$. In any representation of a problem,   one uses, within a certain framework, certain mathematical structures. These are usually implicit in the derivation of the no-go result. We will concentrate on these mathematical structures for the rest of this section. More specifically, we are interested in how one may understand what $\neg M$ implies methodologically in these cases. For that purpose we need to understand what the relation between the physical situation of interest is and the mathematical structure representing it. We will not be concerned with the details of the semantics of physical theories, though relevant, but take a more pragmatic attitude of the  relationship between the mathematical structure and the physical situation. 

Let us again consider the mathematical representation of symmetries. Symmetries are usually represented in terms of the algebraic structure of groups. There are different kinds of groups for different kinds of symmetries. 
In order to understand the implications of  $\neg M$, we need to address the uniqueness of the mathematical representation concerning the physical phenomenon of interest. First, one may ask whether there is only a unique group able to represent the situation of interest. This is usually not the case and has been discussed in the literature on structural underdetermination. \citet{roberts2011group} for instance, posing it as a problem for supporters of group structural realism, shows how one can understand a group $\mathbb{G}$ as well as its automorphism group $Aut(\mathbb{G})$ as a basis from which one can construct the physical situation.\footnote{This is not true for all groups as some groups, e.g. the permutation group $S_3$, is isomorphic to its automorphism group. See  \citet[p.62]{roberts2011group} for more details.} This, of course, goes on including the automorphism group of the automorphism group of $\mathbb{G}$ and so on. So there is a whole `hierarchy' of symmetry groups one can consider in representing the physical situation. 

Second, one may consider whether groups are the unique structure able to represent the situation.
 Both $\mathbb{G}$ and $Aut(\mathbb{G})$, although different groups, are still the same algebraic structure, in the sense that they both satisfy the same algebraic axioms, namely those of groups. There are, however, many algebraic structures we could in principle use to represent  symmetries. As we saw in the case of supersymmetry, it was exactly this move from one algebraic structure, namely Lie algebras, to another algebraic structure, namely $\mathbb{Z}_2$-graded Lie algebras, that allowed internal and external symmetries to combine non-trivially.  Graded Lie algebras can be understood as generalisations of Lie algebras. In this sense, everything a Lie algebra can describe can also be described by a graded Lie algebra; the converse however is not true. This kind of generalisation is usually a possible methodological option. 
 
 Consider the  requirement that probabilities satisfy the Kolmogorov axioms. As we saw in the previous section, certain quantum mechanical probabilities violate the axioms of Kolmogorov. We do not want to say that they are therefore not probabilities but instead that they may satisfy different axioms of probability, i.e. they are  non-Kolmogorovian probabilities.  If we want to change the structure, we can consider weakening one of the axioms, e.g. the additivity axiom, leading to what is sometimes called upper or lower probability spaces. This will similarly count as a more generalised structure in the sense that all Kolmogorovian probabilities will satisfy these changed axioms as well. Another option, however, would have been to allow for negative probabilities. This again would still allow us to account for all Kolmogorovian probabilities. These other non-Kolmogorovian probabilities can still be affected by \emph{other} no-go theorems, but provide, at first, methodological options in need of further analysis.\footnote{No-go theorems for further non-Kolmogorovian probabilistic approaches to quantum mechanics exist. See e.g.  \citep{feintzeig2017noncontextual}.}
 
 So to sum up, the $\neg M$ route opens up different strategic options. The argument for or against a specific choice of mathematical structure  is usually in need of an independent evaluation. For instance, one may argue based on simplicity arguments in favor of one structure being more fundamental than another.\footnote{See e.g. \citet{north2009structure} for an argument along that line in favor of the structure associated with the Hamiltonian formulation of classical mechanics} These arguments heavily depend on the kind of simplicity measures used and arguments for either one may be lacking \cite[p. 303]{curiel2014classical}. Similarly, one may argue against certain non-Kolmogorovian probabilities by the lack of suitable interpretations for them. All of these are independent justifications one may give for a certain mathematical structure over the others that need further elucidation, usually pointing to further underlying assumptions.

\section{No-Go Theorems: What Are They Good For?}\label{whereto}

No-go theorems are complicated and  hard to dissect. We have provided a possible abstract definition of no-go theorems, which allowed us to analyze it in more detail and to comprehend it in a more fine-grained way. We would now like to draw some more general conclusions by stating five broad methodological lessons, which are supposed to be complemented by the more detailed analyses of Section 4:

\textbf{Lesson 1}: No-go theorems have a more complex structure than is usually explicitly stated. \newline
The  cases from the history of physics we considered, showed that the often multi-layered structure of no-go theorems does not allow for a straightforward conclusion to be drawn from the theorem by itself.
 As discussed, they are usually posed as either impossibility results with respect to the goal G or some element of the physical assumptions $P$. This simplified picture ignores the important role played by the framework $F$ and the mathematical structure $M$ and the strategic options they offer.

\textbf{Lesson 2}: The no-go theorem itself does not state which element of the theorem to give up. \newline  A no-go theorem is a contradiction, which derives from a set of elements. The result itself does not say, which of the elements involved in the derivation is more  and which one is less justified and so does not entail the rejection of any one specific element. 

\textbf{Lesson 3}: There is not a unique implication one can draw from no-go theorems.
\newline
This is a corollary from the previous lesson. Once we have established a no-go theorem we need to address the question, how we wish to address the contradiction, i.e. how we wish to interpret the no-go result. The interpretation depends on which of the elements of the no-go theorem we are most willing to change or give up on. However, as we have seen, not all elements are empirical certainties of nature, but vary strongly based on the justifications one may give for them. Furthermore, different scientists may have different justifications for the elements of a no-go theorem, corresponding to a difference in ordering of what one prefers to give up or change first. This difference in preference assignment will correspond to differences in interpreting the same no-go theorem. So it is important to recognize that there is not a  unique implication one can draw from a no-go theorem by itself. 

\textbf{Lesson 4}: The consideration of the mathematical structure $M$ deserves more recognition. \newline In principle, we can imagine an empirically motivated goal $G$ and similarly empirically well-confirmed physical assumptions $P$ within a determined framework $F$. We cannot claim the same for mathematical structures. While one may be committed to a certain goal  and physical assumptions, this is usually less so with the mathematical structures.   We may have many good reasons to choose one mathematical structure rather than another, based on simplicity and naturalness assumptions. But the empirical access to them is very limited. Keeping certain physical assumptions fixed one can empirically only point to the insufficiency of a certain mathematical structure to account for some observed phenomenon. This leaves a whole lot of weaker and therefore more encompassing structures untouched. The space of all mathematical structures is not a clearly defined space\footnote{This amplifies the previous lesson, by drawing attention to the imprecise space that is being opened up by the no-go theorem.}. As such, it does not allow for a rigorous ``working through all structures''-approach, but only allows for theoretical exploration. This naturally leads to the methodological implications of no-go theorems, which comes in our next lesson.


\textbf{Lesson 5}: No-go theorems are  (at first) best understood as  go theorems. \newline No-go theorems usually  do not strictly speaking allow for an interpretation  as an impossibility result with respect to some $G$ or $P$, as that would imply one has certainty with respect to the rest of the elements and this is, as we saw above, usually not  the case. 
 So what do they imply? If we, for instance, accept the mathematical structure $M$ as the ``weakest'' element, i.e. the element we are least committed to, we interpret the no-go theorem as implying $\neg M$. But as we have already said, $\neg M$ cannot meaningfully be interpreted as the impossibility of the mathematical structure, but as an invitation to consider alternative mathematical structures to replace it. This may lead to new no-go theorems (as discussed above) or to unacceptable physical consequences in which case one obtains support for the original assumption $M$. This would strengthen the impossibility interpretation of the theorem. Alternatively, a new mathematical structure $M'$ may be able to circumvent the initial no-go theorem without leading to physical problems (as the replacement of Lie algebras by graded Lie algebras allowed for the non-trivial combination of internal and external symmetries). Before the exploration of $\neg M$ one simply does not know. It is in this sense that no-go theorems are at first best understood as go-theorems, i.e. as outlining the possible methodological pathways in pursuing to show the possibility or impossibility of some goal $G$. 
 They are excellent tools in theory development, while being (at first) unreliable tools in stopping research programmes.

\section{Conclusion}
We started with a case study of the development of a no-go theorem from particle physics, which provided us with enough detail to recognise the different abstract elements of  no-go theorems. We discussed each element in detail coming to the conclusion that no-go theorems cannot at first be understood as impossibility results in the strict sense. Especially, the mathematical structure $M$ poses a threat to this strong conclusion. This turned the role of no-go theorems around. Rather than understanding a no-go theorem as providing us  direct insights into what is  not possible in the world, they should be understood as a methodological starting point in theory development, where in the end we may be able to circumvent it or become more and more certain that we are less willing to give up certain assumptions to make something possible. 



While we have outlined a more systematic analysis of no-go theorems, we could have chosen an alternative route to the same conclusion, namely via meta-induction on the history of physics. Von-Neumann's no-go theorem was superseded by both, actual hidden variable theories (pilot wave theories, Bohmian mechanics), and further no-go theorems where  the physical assumptions $P$, the framework $F$ as well as the  mathematical structures $M$ have been changed. The impossibility to simulate chiral fermions on a lattice, the Nielsen-Ninomiya no-go theorem, was circumvented  via the introduction of domain wall fermions  by extending the mathematical representation of the lattice with an additional dimension \citep{kaplan1992method,shamir1993chiral}. \citet{weinberg1980limits}
proved that gravitons cannot be composite particles in a relativistic quantum field theory. There is now a whole plethora of counter examples: from conformal field theories and massive gravity to String theory\footnote{See \citep{bekaert2012higher} for a review article on how the Weinberg-Witten theorem is circumvented in these theories.}. We have already discussed Supersymmetry and how it circumvented the Coleman-Mandula theorem by a change in $M$. We could continue with other examples, but this should suffice for our purposes. One could now argue, based on this historical evidence, that maybe current no-go theorems will be superseded by ways to circumvent them as well. This is in complete agreement with our analysis above. That is, it was to be expected that no-go theorems do not say the last word with respect to one's goal $G$. Our analysis actually provides the explanation why they do not. However, history is also full of examples where these no-go theorems did actually have the effect of stopping whole research programmes. That is, we have many historical examples where no-go theorems were systematically misunderstood in what they can imply.  
So no-go theorems have played a role in the history and methodology of physics, for which they did not provide the argumentative support. There is a discrepancy  between what no-go theorems \emph{can} imply and how they were actually interpreted in practice. Recognising what they can imply provides us with a more adequate use of them as a tool in theory development. This more adequate use is the understanding that no-go's are (at first) actually the best go's!

\section*{Acknowledgements}
This work was supported by the DFG (grant FOR 2063).
I would like to thank   Alexander Blum, Richard Dawid, Juliusz Doboszewski, Stephan Hartmann, Niels Linnemann, Owen Maroney and Karim Thebault  for various discussions on the topic of this paper. I am grateful to two anonymous referees for helping clarify several points.  Special thanks to Mike Cuffaro, Erik Curiel and Gregor Schiemann for detailed comments on an earlier draft.


\bibliographystyle{ChicagoReedWeb}

\end{document}